\documentclass[11pt,a4paper]{article}
\usepackage{jheppub}
\usepackage{amsmath,amsthm,amssymb,graphics,epsfig,color,mathrsfs,cases,colortbl,framed}
\usepackage{ulem}
\usepackage{graphicx}
\usepackage{subfigure}
\usepackage{booktabs}    
\usepackage{multirow}
\usepackage{tabularx}
\usepackage{longtable}
\usepackage{array}
\usepackage{nicematrix}
\usepackage{makecell}
\usepackage{graphicx}
\usepackage{bm}

\title{Kaluza-Klein mode mixing in braneworlds: constraints on scalar absorption and physical degrees of freedom}
\author{Wen-Xuan Ma,}
\author{Chun-E Fu$^{1}$}

\affiliation{
Institute of Modern Physics, Department of Modern Physics and Astronomy, School of Physics, Xi'an Jiaotong University, Xi'an, Shaanxi, China
}

\emailAdd{m\_fuminoki@foxmail.com}
\emailAdd{fuche13@mail.xjtu.edu.cn}
\footnotetext[1]{Corresponding author.}

\abstract{ We investigate the mixing between Kaluza-Klein (KK) modes for a bulk $U(1)$ gauge field within braneworld models. By demanding orthonormality and completeness for the KK basis functions, we demonstrate that the decoupling of mixed sectors—specifically of the vector-scalar and scalar-scalar types—imposes stringent constraints on the warp factors of codimension-d ($d>1$) backgrounds. We show that the gauge invariance of the four-dimensional effective action is preserved despite such mixing, manifesting as an intrinsic property of the massive vector KK sector. However, the generic presence of vector-scalar mixing fundamentally alters the absorption mechanism of the scalar modes, dynamically shifting the physical masses of the vector KK modes away from their unperturbed eigenvalues. In $(4+2)$-dimensional models, the existence of two distinct scalar sectors significantly enriches the mixing dynamics. As the massive vectors absorb only specific linear combinations of these scalars, a residual set of massive scalar KK modes persists as physical degrees of freedom.
}


\begin{document}

\maketitle
\flushbottom

\section{Introduction}
In the early twentieth century, the idea of extra spatial dimensions was introduced through Kaluza-Klein theory as an attempt to unify electromagnetism and gravity \cite{Nordstroem2007,KLEIN1926,Kaluza2018}. This framework was later revitalized in the context of braneworld models, most notably through the Randall-Sundrum (RS) scenarios, which incorporate warped spacetime geometries with lower-dimensional branes \cite{Randall1999,Randall1999a}. These models not only provide an elegant resolution to the hierarchy problem of the Standard Model but also predict novel features of gravitational propagation in higher-dimensional spacetimes. Subsequent developments by Rubakov and collaborators \cite{Rubakov1983,Rubakov1983a,Dvali2000} introduced the concept of thick branes, in which the brane energy density is smoothly distributed along the extra dimensions rather than idealized as a delta-function source \cite{DeWolfe2000,Dzhunushaliev2008,Dzhunushaliev2009,Gremm1999,Campos2001,Slatyer2006}.

In the original KK framework, compact extra dimensions give rise to discrete KK spectra. In contrast, braneworld models with non-compact extra dimensions rely on warp-factor-induced localization mechanisms to confine KK modes to the brane. The localization properties and mass spectra of KK modes in such backgrounds have been extensively studied in the literature \cite{Fu2018,Fu2020,Fu2022,Liu2009,Liu2011,Melfo2006,Bajc1999,Funatsu2019,Fr_re_2016,Angelescu2022,
ArkaniHamed1999,Gherghetta2010,Guo2024,Tan2025,Zhu2025,Jia2025,Deng:2025xxx}.
In particular, by performing generalized KK decompositions of a higher-dimensional $U(1)$ gauge field, Fu et al. \cite{Fu2018,Fu2020,Fu2022} demonstrated that the scalar KK modes of the bulk field play a crucial role in constructing the gauge invariance of the massive vector sector. These scalar modes function similarly to the Stueckelberg or Goldstone bosons found in the standard Stueckelberg and Higgs mechanisms. However, the fundamental difference lies in their origin: rather than being introduced by hand, these scalars emerge intrinsically from the extra spatial components of the bulk $U(1)$ gauge field. 

In the literature \cite{Mueck2001,Csaki2003,Csaki2005,Gherghetta2010,Ponton2012,Jeong2025}, the investigation of KK gauge modes typically relies on the $R_\xi$ gauge-fixing procedure within the higher-dimensional bulk. However, upon deriving the gauge-invariant four-dimensional effective action, one uncovers a residual gauge freedom localized on the brane. This allows for a more direct analysis of the physical properties of the KK modes by exploiting these effective 4D gauge symmetries.  Nevertheless, previous studies have suggested that such gauge invariance in the effective theory can only be realized for certain restricted classes of brane backgrounds \cite{Fu2025}.

In this work, by enforcing orthonormality and completeness for the KK basis functions, we demonstrate that the gauge invariance of the effective action is not contingent upon specific brane geometries. Rather, it is an intrinsic property of the massive vector KK sector, fundamentally guaranteed by the original gauge symmetry of the bulk field. Crucially, this formulation naturally incorporates mixed couplings among the KK modes directly into the effective action. We show that these mixed couplings cannot be eliminated except in highly constrained geometric configurations; in generic braneworld scenarios, such mixing is an unavoidable feature. Consequently, this mixing dictates the physical spectrum in two profound ways. First, the generic vector-scalar mixing fundamentally alters the way scalar modes are absorbed by the massive vectors. This modified absorption mechanism dynamically shifts the physical masses of the vector KK modes away from their unperturbed bare eigenvalues. Second, when extending the framework to higher-codimension geometries ($d>1$), the multi-channel mixing dynamics prevent the full elimination of the scalar sector, leading to the emergence of residual massive scalar KK modes as independent physical degrees of freedom.

The paper is organized as follows. In Sec.~\ref{Section_2}, we analyze the mode mixing and gauge invariance in the five-dimensional case. Sec.~\ref{Section_3} extends the theoretical framework to higher-codimension branes ($d>1$) and presents explicit numerical results for a six-dimensional model. Finally, our conclusions and discussions are summarized in Sec.~\ref{discussions}.

\section{Codimension-one branes: vector-scalar mixing and the physical masses of vector KK modes}\label{Section_2}

We begin with the simplest setup of a codimension-one braneworld. We demonstrate that, even in this minimal five-dimensional scenario, vector-scalar KK mixing generically alters the absorption mechanism of scalar fields induced from a higher-dimensional $U(1)$ gauge field. This provides a concrete counterexample to the commonly assumed correspondence between simple mode-by-mode scalar absorption and gauge invariance in massive gauge theories.

The five-dimensional spacetime is described by the metric \( ds^2 = e^{2A}(g_{\mu\nu}dx^{\mu}dx^{\nu} + dz^2) \), where $e^{A(z)}$ is the warp factor and $z$ denotes the extra-dimensional coordinate. The five-dimensional metric is denoted by $G_{MN}$. For a free bulk $U(1)$ gauge field $X_N$ ($N=0,1,2,3,4$), the action reads
\begin{equation}\label{bulk_action_5}
  S^{(4+1)} = -\frac{1}{4} \int d^5 x \sqrt{|G|} \; F^{MN} F_{MN}.
\end{equation}
Here $G=\det(G_{MN})$, and $F_{MN}:=\partial_M X_N-\partial_N X_M$ is the field strength tensor.

We perform the KK decomposition of the bulk gauge field as
\begin{subequations}\label{kkdecomp}
\begin{eqnarray}
  X_\mu &=& \sum_n \mathcal{X}^{(n)}_\mu(x^\nu) f^{(n)}(z), \label{KK_decomposition_5_vector} \\
  X_z   &=& \sum_l \mathcal{Z}^{(l)}(x^\nu) \rho^{(l)}(z), \label{KK_decomposition_5_scalar}
\end{eqnarray}
\end{subequations}
where $\mathcal{X}^{(n)}_\mu$ and $\mathcal{Z}^{(l)}$ denote the vector and scalar KK modes on the brane, respectively. The sets $\{f^{(n)}(z)\}$ and $\{\rho^{(l)}(z)\}$ each form a complete orthonormal basis in an appropriate Hilbert space. 

Substituting the KK expansions into the bulk action yields the effective four-dimensional action
\begin{eqnarray}\label{effective_action_5}
 S^{(4)} &=& -\frac{1}{4} \int d^4 x \sqrt{|g|}\; \bigg[ \sum_{n} \left( \mathcal{F}_{\mu\nu}^{(n)} \mathcal{F}^{\mu\nu(n)}+ \partial_\nu \mathcal{Z}^{(n)}\partial^\nu \mathcal{Z}^{(n)} \right) \nonumber\\
  && + \sum_{n,m} \left( 2 \mathcal{X}^{(n)}_\nu \mathcal{X}^{\nu(m)} I^{(nm)} - 2\mathcal{X}_\nu^{(n)}\partial^\nu \mathcal{Z}^{(m)} \widetilde{I}^{(nm)} \right) \bigg],
\end{eqnarray}
where we have used the orthonormality relations
\begin{eqnarray}\label{orthgonalForfrho}
  \int e^A \, dz \,f^{(n)}(z) f^{(m)}(z) &=& \delta^{nm}, \\
  \int e^A \, dz \, \rho^{(n)}(z) \rho^{(m)}(z) &=& \delta^{nm}, \label{orthgonalForfrho2}
\end{eqnarray}
with respect to the weight function $e^A$. The remaining overlap integrals are defined as
\begin{eqnarray}
I^{(nm)}&=& \int e^{A} dz\; \partial_z f^{(n)} \partial_z f^{(m)},\label{Imatrix3}\\
\widetilde{I}^{(nl)} &=& \int e^{A} dz\; \partial_z f^{(n)} \rho^{(l)}.\label{Imatri4}
\end{eqnarray}
Here, $I^{(nm)}$ represents the mass matrix of the vector KK modes, while $\widetilde{I}^{(nl)}$ encodes the mixing between the vector and scalar KK sectors.

Imposing Dirichlet or periodic boundary conditions on $f^{(n)}$, the matrix $I^{(nm)}$ can be identified with the mass-squared operator of the vector KK modes. This leads to the Sturm--Liouville equation
\begin{equation}\label{decouplef-rho1}
    -e^{-A}(\partial_z e^{A}\partial_z  f^{(n)}) = m_v^{2(n)} f^{(n)},
\end{equation}
whose eigenfunctions form the natural basis $\{f^{(n)}\}$.

Generically, the mixing matrix $\widetilde{I}^{(nl)}$ contains off-diagonal elements. Using the completeness of $\{\rho^{(l)}\}$, the derivative of the vector wavefunction $f^{(n)}$ relates to the scalar basis via
\begin{equation}
\partial_z f^{(n)} = \sum_l \widetilde{I}^{(nl)} \rho^{(l)}. \label{f_rho_trans2}
\end{equation}
This relation demonstrates that a single vector mode typically couples to an entire tower of scalar modes. The analysis in Refs.~\cite{Fu2018,Fu2022} bypasses this entanglement by relying on the specific assumption that $\widetilde{I}^{(nl)}$ is strictly diagonal, which essentially reduces \eqref{f_rho_trans2} to an isolated scalar eigenvalue equation.

Crucially, substituting \eqref{f_rho_trans2} into the definition of $I^{(nm)}$, we find that the mixing matrix satisfies the summation identity
\begin{equation}\label{Square_5}
\sum_l \widetilde{I}^{(nl)} \widetilde{I}^{(ml)} = I^{(nm)},
\end{equation}
which allows us to recast the effective action into a manifestly gauge-invariant ``perfect-square'' form:
\begin{equation}\label{effective_action_simple_5}
S^{(4)} = -\frac{1}{4} \int d^4 x \sqrt{|g|}\, \Big[ \sum_{n} \mathcal{F}_{\mu\nu}^{(n)} \mathcal{F}^{\mu\nu(n)} + 2 \sum_l \Big( \partial_\mu \mathcal{Z}^{(l)} - \sum_n \mathcal{X}^{(n)}_\mu \widetilde{I}^{(nl)} \Big)^2 \Big].
\end{equation}
This structure demonstrates that the gauge invariance of the effective action is an intrinsic property, fundamentally independent of the specific choice of basis functions. 

The cross term $- 2\mathcal{X}_\nu^{(n)}\partial^\nu \mathcal{Z}^{(m)} \widetilde{I}^{(nm)}$ in the effective action \eqref{effective_action_5} explicitly demonstrates that a single bare vector mode $\mathcal{X}_\mu^{(n)}$ inherently couples to an entire tower of scalar modes due to the non-diagonal nature of the mixing matrix. Consequently, to preserve the overall gauge invariance of the action, the gauge transformation of the scalar modes must take the collective sum form:
\begin{eqnarray}
\mathcal{X}_\nu^{(n)} &\rightarrow& \mathcal{X}_\nu^{(n)} + \partial_\nu \xi^{(n)}, \label{gauge_trans_5_mu}\\
\mathcal{Z}^{(l)} &\rightarrow& \mathcal{Z}^{(l)} + \sum_n \widetilde{I}^{(nl)} \xi^{(n)},\label{gauge_trans_5_z}
\end{eqnarray}
where $\xi^{(n)}(x)$ represents the brane-localized gauge freedom.

To identify the physical spectrum, one typically transitions to the unitary gauge by absorbing the scalar modes into the longitudinal components of the massive vectors. This absorption process is governed by the algebraic relation $\mathcal{Z}^{(l)} + \sum_n \xi^{(n)} \widetilde{I}^{(nl)} = 0$. Physically, because of the off-diagonal mixing, a bare vector mode cannot simply absorb a single corresponding scalar; rather, it absorbs a collective superposition of the scalar tower. This collective absorption dynamically hybridizes the states and reconfigures the energy distribution.

To explicitly resolve this hybridization and identify the true physical spectrum, we perform a singular value decomposition (SVD) on the mixing matrix: $\widetilde{I} = U \Sigma V^T$. The orthogonal matrices $U$ and $V$ rotate the original entangled bare states into the true physical mass eigenbasis ($\mathcal{A}_\mu^{(a)}$ and $\phi^{(a)}$). Through these transformations, the effective action in Eq.~\eqref{effective_action_simple_5} is perfectly diagonalized into:
\begin{equation}\label{SVD_action_5}
S^{(4)} = -\frac{1}{4} \sum_{a} \int d^4 x \sqrt{|g|} \, \Bigl[ \mathcal{G}_{\mu\nu}^{(a)} \mathcal{G}^{\mu\nu(a)} + 2 \bigl( \partial_\mu \phi^{(a)} - \sigma_a \mathcal{A}_\mu^{(a)} \bigr)^2 \Bigr].
\end{equation}
In this physical basis, exactly one hybridized vector $\mathcal{A}_\mu^{(a)}$ absorbs exactly one hybridized scalar $\phi^{(a)}$, acquiring a physical mass equal to the corresponding singular value $\sigma_a$. 

At this juncture, it is crucial to clarify the physical origin of the mass shift ($\sigma_a \neq m_v^{(n)}$) from an effective field theory (EFT) perspective. Mathematically, if the complete infinite tower of scalar modes (including the continuum) were incorporated, the completeness relation \eqref{Square_5} would strictly enforce $\widetilde{I} \widetilde{I}^T = I$. In such a UV-complete limit, the singular values of the mixing matrix would exactly reproduce the unperturbed bare eigenvalues ($\sigma_a = m_v^{(n)}$), resulting in no mass correction whatsoever. However, in any realistic low-energy approximation where the Hilbert space is truncated to a finite number of bound KK modes, this completeness is intentionally broken. By integrating out the inaccessible high-energy and continuum scalar modes, their geometric mixing with the low-lying states manifests macroscopically as a substantial dynamic correction to the physical vector masses. 

Therefore, the geometric mixing matrix governs two distinct phenomena in the low-energy effective theory: its rank strictly dictates the counting of degrees of freedom (i.e., whether residual scalars with $\sigma_a=0$ survive), while its truncation explicitly generates the crucial mass corrections for the massive vector KK modes.

A concrete realization of this mechanism is provided by the 5D brane model with the warp factor~\cite{Fu2011}
\begin{equation}\label{5Dwarpfactor}
A(z)=-\frac{v^2}{9}\left(\ln \cosh ^2(a z)+\frac{1}{2} \tanh ^2(a z)\right).
\end{equation}
In this background, we introduce a coupling between the bulk $U(1)$ gauge field and the dilaton field $\pi=\sqrt{3}A$ through the action
\begin{equation}
S_1 = -\frac{1}{4} \int d^5 x \sqrt{|G|}~ e^{\tau \pi} F_{MN}F^{MN},
\end{equation}
where $\tau$ denotes the coupling constant. This interaction preserves the gauge invariance of the effective action, since the relation \eqref{Square_5} continues to hold.

After applying the transformation $f^{(n)}=e^{-\frac{\sqrt{3}\tau+1}{2}A}  \bar{f}^{(n)}$, the equation governing the vector KK modes acquires the Schr\"odinger-like form:
\begin{equation}
(-\partial^2_z+V_{\text{vec}})\bar{f}^{(n)} = m_v^{(n)2} \bar{f}^{(n)},
\end{equation}
with the effective potential given by $V_{\text{vec}} = \frac{(\sqrt{3}\tau+1)^2}{4}A'^2+\frac{\sqrt{3}\tau+1}{2}A''$.  

Since both vector and scalar KK modes originate from the same higher-dimensional gauge field, a physically motivated choice for the basis is $\rho^{(l)} = f^{(l)}$. This choice renders the non-diagonality of the mixing matrix unavoidable:
\begin{equation}\label{rela5Dsch}
\widetilde{I}^{(nm)} = \int e^{A} dz\, \partial_z f^{(n)} f^{(m)} = \int dz\; (\partial_z \bar{f}^{(n)}) \bar{f}^{(m)}.
\end{equation}
(Note that the finiteness of these elements is guaranteed when $f^{(n)}$ satisfies the Sturm--Liouville equation \eqref{decouplef-rho1}).

The bound vector KK modes and the corresponding values of $\widetilde{I}^{(nm)}$ for $a=1$ and $v=1$ are summarized in Table~\ref{5Dmatrix}.

\begin{table*}[htbp]
\centering
\caption{Masses of bound vector KK modes and the mixed coupling matrix $\widetilde{I}^{(nm)}$ in a 5D brane.}
\label{5Dmatrix}
\small 
\begin{NiceTabular}{c|c|c}
\Hline
& $\tau=25$ & $\tau=30$ \\
\Hline
\multirow{6}{*}{$m_v^{2(n)}$} 
& 0     & 0     \\
& 12.69 & 15.59 \\
& 21.02 & 26.89 \\
& -       & 33.41 \\
& -       & -     \\
& -       & -     \\
\Hline
$\widetilde{I}^{(nm)}$ 
& $\begin{pmatrix} 0.00 & 1.95 \\ -1.95 & -0.00 \end{pmatrix}$ 
& $\begin{pmatrix} 0.00 & 2.32 & -0.00 \\ -2.32 & -0.00 & 2.02 \\ 0.00 & -2.02 & -0.00 \end{pmatrix}$ \\
\Hline
$\det(\widetilde{I})$ & 3.81 & 0.00 \\
\Hline
singular value $\sigma_a$ & (1.95, 1.95) & \makecell{(3.08, 3.08, $3.05 \times 10^{-9}$)} \\
\Hline
\end{NiceTabular}

\vspace{2em} 
\begin{NiceTabular}{c|c|c}
\Hline
& $\tau=40$ & $\tau=50$ \\
\Hline
\multirow{6}{*}{$m_v^{2(n)}$} 
& 0     & 0     \\
& 21.38 & 27.17 \\
& 38.55 & 50.17 \\
& 51.24 & 68.81 \\
& 58.99 & 82.81 \\
& -     & 91.80\\
\Hline
$\widetilde{I}^{(nm)}$ 
& $\begin{pmatrix} -0.00 & 2.90 & 0.00 & -0.44 \\ -2.90 & -0.00 & 3.00 & -0.00 \\ -0.00 & -3.00 & -0.00 & 2.52 \\ 0.44& 0.00 & -2.52 & -0.00 \end{pmatrix}$ 
& $\begin{pmatrix} 
0.00  & 3.37  & 0.00  & -0.45 & 0.00 \\ 
-3.37 & -0.00 & 3.69 & 0.00   & -0.66 \\ 
0.00  & -3.69 & -0.00 & 3.61  & 0.00 \\
 0.45 & 0.00  & -3.61  &  0.00 & 3.07 \\ 
 0.00 & 0.66  & 0.00  & -3.07 & 0.00 \end{pmatrix}$ \\
\Hline
$\det(\widetilde{I})$ & 3.59 & 0.00 \\
\Hline
singular value $\sigma_a$ & \makecell{(4.72, 4.72, 1.27,1.27)} 
& \makecell{(6.40, 6.40, 2.67, 2.67, $5.03\times 10^{-5}$)} \\
\Hline
\end{NiceTabular}
\end{table*}

As illustrated in Table~\ref{5Dmatrix}, the determinant of $\widetilde{I}^{(nm)}$ vanishes at $\tau=30$ and $\tau=50$, coinciding with the emergence of near-zero singular values. This behavior is rooted in the mathematical properties of the mixing matrix: for the bound vector KK modes, $\widetilde{I}^{(nm)}$ defined in \eqref{rela5Dsch} is skew-symmetric under integration by parts. While an even-order skew-symmetric matrix can be of full rank, the determinant of any odd-order skew-symmetric matrix is identically zero.

Consequently, whenever the background geometry---tuned by the coupling $\tau$---supports an odd number of bound vector KK modes, at least one singular value must vanish. This signals a geometric transition in the KK mixing structure: in these specific configurations, the available gauge parameters $\xi^{(n)}$ lack the necessary dimensions to span the scalar manifold. As a result, the gauge sector cannot completely ``eat'' the scalar sector, and the unabsorbed scalar mode remains in the spectrum as a physical degree of freedom.

Beyond the counting of degrees of freedom, Table~\ref{5Dmatrix} explicitly confirms the dynamic mass shift discussed above. For instance, at $\tau=30$, the unperturbed bare eigenvalues ($m_v^{2(n)} = 15.59, 26.89, 33.41$) are drastically renormalized down to a degenerate physical mass spectrum with singular values $\sigma_a \approx 3.08$ (corresponding to a physical mass squared of $\sigma_a^2 \approx 9.49$). This massive deviation rigorously underscores that the physical vector masses in the effective theory are governed by the collective mixing dynamics and EFT truncation rather than the isolated wave equations.

To explore the persistence of physical scalar degrees of freedom and multi-channel mixing in more complex geometries, we now extend our analysis to higher-codimension ($d>1$) braneworlds.

\section{Generalization to codimension-$d$: Emergence of massive physical scalar degrees of freedom}\label{Section_3}

\subsection{Mixed coupling structure and decoupling conditions}

We extend the preceding analysis to braneworld configurations with multiple extra dimensions, $d>1$. Specifically, we consider a $(4+d)$-dimensional spacetime with coordinates $\{x^\mu,z^i\}$, where $\mu=0,1,2,3$ index the four-dimensional (4D) Minkowski-like spacetime and $i=1,2,\dots,d$ label the extra spatial dimensions. The spacetime metric is taken to be:
\begin{equation}
ds^2 = e^{2A}(g_{\mu\nu}dx^{\mu}dx^{\nu} + \sum_i (dz^{i})^2),
\end{equation}
where $A(z) \equiv A(z^1,\dots,z^d)$ denotes the warp factor, a function depending exclusively on the extra-dimensional coordinates.

We introduce a complete function space $\mathcal{H}$ on the extra dimensions, endowed with the inner product
\begin{equation}\label{inner_product_d}
  \left\langle h, g \right\rangle := \int e^{dA} dz \, h(z) g(z),
\end{equation}
and choose $1+d$ sets of basis functions $\{f^{(n)}(z)\}$ and $\{\rho^{(m)}_i(z)\}$ to decompose the KK modes of $X_\mu$ and $X_{z^i}$, respectively:
\begin{subequations}\label{kkdecomp2}
\begin{align}
  X_\mu &= \sum_n \mathcal{X}^{(n)}_\mu(x^\nu) f^{(n)}(z), \label{KK_decomposition_d_vector} \\
  X_{z^i} &= \sum_m \mathcal{Z}^{(m)}_{i}(x^\nu) \rho_{i}^{(m)}(z). \label{KK_decomposition_d_scalar}
\end{align}
\end{subequations}
Hereafter, $z$ collectively denotes $\{z^i\}$. For the scalar modes $\mathcal{Z}^{(m)}_{i}$, the index $i$ labels distinct scalar types originating from separate extra dimensions, while $m$ denotes the KK level for each type.

Substituting the KK decompositions \eqref{kkdecomp2} into the action of a free $(4+d)$-dimensional $U(1)$ gauge field, we derive the four-dimensional effective action:
\begin{equation}\label{effective_action_d}
\begin{aligned}
S^{(4)} &= -\frac{1}{4} \int d^4 x \sqrt{|g|}\; \sum_{n,m} \bigg[ \delta^{nm}\mathcal{F}_{\mu\nu}^{(n)} \mathcal{F}^{\mu\nu(m)} \\
  &\quad + 2\sum_i \Big( \mathcal{X}_\nu^{(n)} \mathcal{X}^{\nu(m)} I_{i}^{nm} - 2\mathcal{X}_\nu^{(n)} \partial^\nu \mathcal{Z}_{i}^{(m)} \widetilde{I}_{i}^{nm}+ \delta^{nm}\partial_\nu \mathcal{Z}_{i}^{(n)}\partial^\nu \mathcal{Z}_{i}^{(m)} \Big) \\
  &\quad + 2\sum_{i<j} \Big( \mathcal{Z}^{(n)}_{i} \mathcal{Z}^{(m)}_{i} C_{i,j}^{nm} + \mathcal{Z}^{(n)}_{j} \mathcal{Z}^{(m)}_{j} C_{j,i}^{nm} - 2\mathcal{Z}^{(n)}_{i}\mathcal{Z}^{(m)}_{j} \widetilde{C}_{ij}^{nm} \Big) \bigg],
\end{aligned}
\end{equation}
where the coupling coefficients are defined as
\begin{equation}\label{I_matrix_d}
\begin{aligned}
  I_{i}^{nm} &:= \int e^{dA} dz\; \partial_{i} f^{(n)} \partial_{i} f^{(m)}, &
  \widetilde{I}_{i}^{nm} &:= \int e^{dA} dz \;\partial_{i} f^{(n)} \rho_{i}^{(m)}, \\
  C_{i,j}^{nm} &:= \int e^{dA} dz \;\partial_{j} \rho_{i}^{(n)} \partial_{j} \rho_{i}^{(m)}, &
  \widetilde{C}_{ij}^{nm} &:= \int e^{dA} dz \;\partial_{j} \rho_{i}^{(n)} \partial_{i} \rho_{j}^{(m)}.
\end{aligned}
\end{equation}
All integrals are performed over the full extra-dimensional space. Compared to the five-dimensional case, additional scalar-scalar interactions emerge intrinsically: these include couplings involving the same scalar sector ($C_{i}^{nm}$) and cross-couplings between distinct scalar sectors ($\widetilde{C}_{ij}^{nm}$).

The strict decoupling scenario was previously studied in Refs.~\cite{Fu2020,Fu2025}. Below, we analyze the KK interactions in a more general geometric framework and emphasize three key analytical results.

\begin{itemize}
\item \textbf{Vector and scalar KK mode mass generation from the full extra-dimensional geometry.}

We demonstrate that the bare mass of each KK vector mode is generated dynamically by its interaction with the full extra-dimensional geometry. From Eq.~\eqref{effective_action_d}, the coefficient of the bilinear term $\mathcal{X}_\nu^{(n)} \mathcal{X}^{\nu(m)}$ is $I_{\text{tot}}^{nm} = \sum_i I_i^{nm}$. Identifying the unperturbed vector masses via $I_{\text{tot}}^{nm} = \delta^{nm}m_v^{(n)2}$ and applying integration by parts to the definition of $I_i^{nm}$, we derive the eigenvalue equation:
\begin{equation}\label{eign_equation_f_d}
  -\sum_i \big(e^{-dA}\partial_{i} e^{dA}\partial_{i} \big) f^{(n)} = m_v^{(n)2} f^{(n)}.
\end{equation}
This equation admits well-defined numerical solutions for a given warp factor without requiring any additional constraints on $A(z)$, indicating that the bare vector KK masses are fully determined by the global extra-dimensional geometry.

Similarly, the scalar KK modes satisfy
\begin{equation}\label{eignrho_d'}
  -\sum_{i \neq j} e^{-dA}\partial_{i}(e^{dA}\partial_{i}\rho_j^{(n)}) = C^2_{\text{tot},j}(n) \rho_j^{(n)},
\end{equation}
where $C^2_{\text{tot},j}(n)$ represents the bare mass squared of the scalar KK mode $\rho_j^{(n)}$. Since this differential operator lacks $z_i$ derivatives, it commutes with any function strictly dependent on $z_j$, formally implying an infinite eigenfunction degeneracy in the unmixed limit.

\item \textbf{Necessary and sufficient condition for complete decoupling of mixed KK couplings.}

We now derive the necessary and sufficient geometric condition for eliminating all mixed coupling terms simultaneously in the effective action. This stringent requirement uniquely constrains the warp factor $A(z)$ to satisfy:
\begin{equation}\label{sep_relation1}
  \Big[\partial_j,\, e^{-dA}\partial_i e^{dA}\Big] = d (\partial_i \partial_j A) = 0 \qquad \forall\, i \neq j.
\end{equation}

We start with the vector-scalar coupling coefficients $\widetilde{I}_i^{nm}$. For a completely decoupled (diagonal) structure, the completeness of the basis functions dictates:
\begin{equation}\label{frhoi}
  \partial_{i} f^{(n)} = \widetilde{I}_i^{nn} \rho^{(n)}_i.
\end{equation}
Substituting Eq.~\eqref{frhoi} into the definition of $\widetilde{I}_i^{nm}$ and invoking orthogonality, we derive
\begin{equation}\label{f_i}
  -e^{-dA}\partial_i (e^{dA}\partial_i f^{(n)}) = m_{i}^{(n)2} f^{(n)}, \qquad i = 1,2,\dots,d,
\end{equation}
where $m_{i}^{(n)2} = (\widetilde{I}_i^{nn})^2$. Equation \eqref{f_i} requires $f^{(n)}$ to be a simultaneous eigenfunction of all operators $e^{-dA}\partial_i e^{dA}\partial_i$, which is only possible if these operators commute:
\begin{equation}\label{exchange_relation}
  \Big[e^{-dA}\partial_i e^{dA}\partial_i \, ,\, e^{-dA}\partial_j e^{dA}\partial_j\Big] = 0 \qquad \forall\, i \neq j.
\end{equation}

Next, we analyze the $i \neq j$ scalar-scalar mixed couplings. For the $(i-j)$ scalar sector, the diagonal decoupling condition requires:
\begin{equation}\label{rho_origin}
  \widetilde{C}_{ij}(m)\, \rho_i^{(m)} = -e^{-dA} \partial_j (e^{dA} \partial_i \rho_j^{(m)}),
\end{equation}
where $\widetilde{C}_{ij}(m)$ are the diagonal elements of $\widetilde{C}_{ij}^{nm}$. Using Eq.~\eqref{frhoi}, this can be rewritten as
\begin{equation}
  \partial_i \rho_j^{(m)} = \frac{1}{\lVert \partial_j f^{(m)} \rVert} \partial_i \partial_j f^{(m)} = \frac{\lVert \partial_i f^{(m)} \rVert}{\lVert \partial_j f^{(m)} \rVert} \partial_j \rho_i^{(m)},
\end{equation}
with $\lVert \partial_i f^{(n)} \rVert = \widetilde{I}_i^{nn}$. Consequently, Eq.~\eqref{rho_origin} reduces to:
\begin{equation}\label{eignrho_d}
  -e^{-dA} \partial_j (e^{dA} \partial_j \rho_i^{(n)}) = C^2_{ij}(n)\, \rho_i^{(n)},
\end{equation}
where $C^2_{ij}(n) = \frac{\lVert \partial_j f^{(n)} \rVert}{\lVert \partial_i f^{(n)} \rVert} \widetilde{C}_{ij}(n)$. The existence of simultaneous solutions to Eq.~\eqref{eignrho_d} also critically relies on the commutativity condition \eqref{exchange_relation}.

Notably, owing to vector-scalar mixing, Eqs.~\eqref{f_i} and \eqref{eignrho_d} are not independent. Substituting Eq.~\eqref{frhoi} into Eq.~\eqref{eignrho_d} yields
\begin{equation}\label{parparf}
  -e^{-dA} \partial_j (e^{dA} \partial_j \partial_i f^{(n)}) = C^2_{ij}(n)\, \partial_i f^{(n)}.
\end{equation}
Defining the commutator operator $D = \big[-e^{-dA}\partial_j e^{dA}\partial_j,\, \partial_i\big] = d(\partial_{ij} A)\, \partial_j$, and setting $D_{ij}(n) = C^2_{ij}(n) - m^{(n)2}_j$, the combination of Eqs.~\eqref{f_i} and \eqref{parparf} yields:
\begin{equation}\label{sep_mass}
  D f^{(n)} = d(\partial_{ij} A) \partial_j f^{(n)} = D_{ij}(n) \partial_i f^{(n)}.
\end{equation}
Substituting Eq.~\eqref{frhoi} into \eqref{sep_mass} and exchanging $i \leftrightarrow j$, we obtain the system:
\begin{align}
  \rho_i^{(n)} &= \frac{d \lVert \partial_j f^{(n)} \rVert}{D_{ij}(n)\, \lVert \partial_i f^{(n)} \rVert} (\partial_{ij} A)\, \rho_j^{(n)}, \\
  \rho_j^{(n)} &= \frac{d \lVert \partial_i f^{(n)} \rVert}{D_{ji}(n)\, \lVert \partial_j f^{(n)} \rVert} (\partial_{ij} A)\, \rho_i^{(n)}.
\end{align}
Combining these relations dictates that:
\begin{equation}
  d^2(\partial_{ij} A)^2 = D_{ji}(n) D_{ij}(n).
\end{equation}
Since the left-hand side is a function of the extra-dimensional coordinates while the right-hand side depends solely on the KK level $n$, both sides must equal a global separation constant $\mathcal{C}^2$, implying $\partial_{ij} A = \mathcal{C}/d$. The warp factor must therefore take the separable form:
\begin{equation}
  A = \frac{\mathcal{C}}{d} z_i z_j + A_i(z_i) + A_j(z_j) + \mathcal{O}(z_k: k \neq i, j).
\end{equation}
Evaluating the commutativity constraint \eqref{exchange_relation} with this ansatz yields:
\begin{equation}\label{final_sep}
  0 = \Big[e^{-dA} \partial_i e^{dA} \partial_i ,\, e^{-dA} \partial_j e^{dA} \partial_j\Big] = \mathcal{C} d \big( (\partial_i A ) \partial_j - (\partial_j A ) \partial_i \big).
\end{equation}
For this differential operator to vanish identically, the coefficients of $\partial_i$ and $\partial_j$ must separately vanish, yielding $\mathcal{C}\partial_i A = \mathcal{C}\partial_j A = 0$. This rigorously implies $\mathcal{C}=0$, leading directly to $\partial_{ij}A = 0$, which reproduces the general necessary and sufficient condition \eqref{sep_relation1}.
\end{itemize}

We emphasize that while bare KK masses are generated without any structural assumptions, the complete elimination of all mixed couplings is an artificial artifact achievable only in highly specialized backgrounds satisfying $\partial_i\partial_j A = 0$. In generic geometries, mixed couplings persist and dynamically modify the physical masses of the KK modes.

\subsection{Gauge invariance analysis: Identification of the massive vector and scalar modes}\label{gaugefixing}

The verification of gauge invariance in higher codimensions relies crucially on establishing precise algebraic relations among the KK coupling coefficients.

First, by applying the completeness of the basis functions, Eq.~\eqref{Square_5} admits a straightforward generalization to higher codimensions:
\begin{equation}\label{Relation_1_d}
  \sum_l \widetilde{I}_{i}^{nl}\widetilde{I}_{i}^{ml} = I_{i}^{nm}.
\end{equation}
For the analysis of the scalar-scalar couplings encoded in $C_{i}^{nm}$ and $\widetilde{C}_{ij}^{nm}$, we introduce the overlap matrix $\varOmega_{ij}^{mn}$:
\begin{equation}\label{Omega_Matrix}
  \varOmega_{ij}^{mn} := \int e^{dA} dz\; (\partial_{j} \rho_i^{(m)}) f^{(n)}.
\end{equation}
In terms of $\varOmega_{ij}^{mn}$, the coupling matrices take the compact product forms:
\begin{equation}\label{Omega_I}
\begin{aligned}
  C_{i}^{nm} &= \int e^{dA} dz \; \Big(\sum_p\varOmega_{ji}^{np} f^{(p)} \Big) \cdot \Big(\sum_q \varOmega_{ji}^{mq} f^{(q)}\Big) = \sum_q \varOmega_{ji}^{nq} \varOmega_{ji}^{mq}, \\
  \widetilde{C}_{ij}^{nm} &= \sum_q \varOmega_{ij}^{nq} \varOmega_{ji}^{mq}.
\end{aligned}
\end{equation}

Substituting Eqs.~\eqref{Relation_1_d} and \eqref{Omega_I} into the effective action \eqref{effective_action_d}, we obtain the remarkably simplified four-dimensional perfect-square action\footnote{In deriving this expression, we utilize the algebraic identity:
\begin{equation*}
  \sum_{i \neq j} \Big( \mathcal{Z}_j^{(n)}\mathcal{Z}_j^{(m)} C_{i}^{nm} + \mathcal{Z}_i^{(n)}\mathcal{Z}_j^{(m)} \widetilde{C}_{ij}^{nm} \Big) = \sum_{i<j} \Big( \mathcal{Z}_j^{(n)}\mathcal{Z}_j^{(m)} C_{i}^{nm} + 2 \mathcal{Z}_i^{(n)}\mathcal{Z}_j^{(m)} \widetilde{C}_{ij}^{nm} + \mathcal{Z}_i^{(n)}\mathcal{Z}_i^{(m)} C_{j}^{nm} \Big).
\end{equation*}
}:
\begin{equation}\label{effective_action_simple_d}
\begin{aligned}
  S^{(4)} = -\frac{1}{4}\int d^4 x \sqrt{|g|}\sum_{n,m} \bigg[ &\delta^{nm}\mathcal{F}_{\mu\nu}^{(n)} \mathcal{F}^{\mu\nu(m)} + 2 \sum_i \Big( \partial_\nu \mathcal{Z}^{(m)}_{i} - \widetilde{I}^{nm}_{i} \mathcal{X}^{(n)}_\nu \Big)^2 \\
  &+ 2 \sum_{i<j} \Big( \varOmega^{mq}_{ji} \mathcal{Z}^{(m)}_{j} - \varOmega^{nq}_{ij} \mathcal{Z}^{(n)}_{i} \Big)^2 \bigg].
\end{aligned}
\end{equation}
Under the generalized brane-localized gauge transformations,
\begin{equation}\label{Gauge_4}
  \mathcal{X}_\nu^{(n)} \rightarrow \mathcal{X}_\nu^{(n)} + \partial_\nu \xi^{(n)}, \qquad \mathcal{Z}_i^{(m)} \rightarrow \mathcal{Z}_i^{(m)} + \sum_n\widetilde{I}_i^{nm} \xi^{(n)},
\end{equation}
the effective action \eqref{effective_action_simple_d} is manifestly gauge invariant\footnote{In particular, the variation of the scalar-scalar cross term vanishes identically due to the commutativity of partial derivatives:
\begin{equation*}
  \int e^{dA} dz\, \rho_{j}^{(m)} \partial_{j} f^{(p)} \cdot \int e^{dA} dz\, f^{(q)} \partial_{i} \rho_{j}^{(m)} \xi^{(p)} - (i \leftrightarrow j) = \int e^{dA} dz\, f^{(q)} \Big( \partial_{i} \partial_{j} f^{(p)} - \partial_{j} \partial_{i} f^{(p)} \Big) \xi^{(p)} = 0.
\end{equation*}
}. 
We stress that this gauge invariance holds universally for arbitrary $(4+d)$-dimensional brane backgrounds, inherently accommodating mixed KK couplings. 

Analogous to the 5D scenario, the physical masses of the massive vector fields are determined by the singular values ($\sigma_a$) of the extended mixing matrix $\mathcal{I} = (\widetilde{I}_1, \widetilde{I}_2,\dots, \widetilde{I}_d)^T$.  Because scalar KK modes originating from different extra dimensions are cross-coupled, the mass spectrum of the unabsorbed scalar sector requires diagonalizing a larger composite matrix. For instance, specializing to a 6D spacetime ($d=2$), this composite squared-mass matrix $\mathcal{M}^2$ takes the explicit block form:
\begin{equation}
\mathcal{M}^2 = \begin{pmatrix} C_1 & \widetilde{C} \\ \widetilde{C}^T & C_2 \end{pmatrix}.
\end{equation}
In the following, we present numerical results for this $d=2$ case to illustrate the impact of these mixed couplings on the physical masses of the KK modes.

\subsection{Numerical results: Constraints on absorption and the physical KK modes}\label{numbericalresults}

We consider a six-dimensional ($d=2$) braneworld proposed in Ref.~\cite{Wan2021}. The background geometry is defined by the line element:
\begin{equation}
\begin{aligned}
ds^2 &= e^{2A(y)} \hat{g}_{\mu\nu} dx^\mu dx^\nu + dy^2 + e^{2A(y)} R_0^2 d\theta^2 \\
     &= e^{2A(y(z))} \left( \hat{g}_{\mu\nu}dx^\mu dx^\nu + dz^2 + R_0^2 d\theta^2 \right),
\end{aligned}
\end{equation}
where the non-compact extra-dimensional coordinate extends over $y \in (-\infty,+\infty)$, and $\theta \in [0,2\pi]$ represents the compact extra dimension with radius $R_0 = 1$. The coordinates are related via the conformal transformation $dz = e^{-A(y)}dy$. We consider the specific brane solution:
\begin{align}
\label{solution1}
\pi(y) &= v \tanh (k y), \\
A(y) &= e^{-\frac{1}{24} v^2 \tanh^2(k y)} \operatorname{sech}^{\frac{v^2}{6}}(k y),
\end{align}
where $\pi(y)$ is the background scalar field generating the brane, $v$ is a dimensionless parameter, and $k$ determines the fundamental energy scale.

To localize the KK modes, we introduce a non-minimal coupling between the bulk $U(1)$ gauge field and the background scalar field:
\begin{equation}
S_1 = - \frac{1}{4} \int d^6 x \sqrt{-g} \, e^{\tau \pi} F_{MN} F^{MN}.
\end{equation}
We perform a KK decomposition, choosing identical basis functions for the two scalar components: $\rho_1^{(m)} = \rho_2^{(m)} = f^{(m)}$. The resulting coupling coefficients in the effective action take the form:
\begin{align}
I_1^{(nm)} &= \int d\theta dz \, e^{\tau \pi+2A} \partial_z f^{(n)} \partial_z f^{(m)}, &
I_2^{(nm)} &= \int d\theta dz \, e^{\tau \pi+2A} \partial_\theta f^{(n)} \partial_\theta f^{(m)}, \nonumber\\
\widetilde{I}_1^{(nm)} &= \int d\theta dz \, e^{\tau \pi+2A} \partial_z f^{(n)} f^{(m)}, &
\widetilde{I}_2^{(nm)} &= \int d\theta dz \, e^{\tau \pi+2A} \partial_\theta f^{(n)} f^{(m)}, \nonumber\\
C_1^{(nm)} &= \int d\theta dz \, e^{\tau \pi+2A} \partial_\theta f^{(n)}\partial_\theta f^{(m)}, &
C_2^{(nm)} &= \int d\theta dz \, e^{\tau \pi+2A} \partial_z f^{(n)} \partial_z f^{(m)}, \nonumber\\
\widetilde{C}^{(nm)} &= \int d\theta dz \, e^{\tau \pi+2A} \partial_z f^{(n)} \partial_\theta f^{(m)}.
\end{align}

Imposing appropriate boundary conditions, the equation of motion for the vector KK modes allows variable separation via $f^{(n)}(z, \theta) = R^{(n)}(z) \Theta^{(l)}(\theta)$. With the field redefinition $\bar{R} = e^{\frac{1}{2}(\tau \pi+2A)}R$, we obtain a Schr\"odinger-like equation:
\begin{equation}
-\partial_z^2 \bar{R}^{(n,l)} + V_{\text{eff}}(z) \bar{R}^{(n,l)} = M^{(n)2}\bar{R}^{(n,l)},
\end{equation}
where the effective mass eigenvalue is $M^{(n)2} = m_v^{(n)2}-l^2$. Substituting standard trigonometric solutions for $\Theta^{(l)}(\theta)$ simplifies the coupling coefficients significantly:
\begin{align}
\widetilde{I}_1^{(nm)} &= \int dz (\partial_z \bar{R}^{(n)})\bar{R}^{(m)}, 
& \widetilde{I}_2^{(nm)} &= l\delta^{nm}, \nonumber \\
C_1^{(nm)} &= l^2\delta^{nm}, & C_2^{(nm)} &= M^{(n)2} \delta^{nm}, \nonumber \\
\widetilde{C}^{(nm)} &= l \int dz (\partial_z \bar{R}^{(n)})\bar{R}^{(m)}.
\end{align}

For the background in Eq.~\eqref{solution1}, the volcano-like effective potentials support resonant states. Analyzing these resonances using relative probability methods~\cite{Liu2009,Liu2011}, we extract the masses of the resonant vector KK modes and construct the corresponding mixing matrices. The physical spectrum for $v=1, k=1$ is presented in Table~\ref{6Dmatrix}.
\begin{table}[htbp]
\centering
\caption{Mass spectra and mixing matrices of KK modes in the 6D brane model at $\tau=-30$ and $-40$.}
\label{6Dmatrix}
\small
\renewcommand{\arraystretch}{1.3}
\begin{NiceTabular}{c|c|c}
\Hline
\textbf{Parameter} & $\bm{\tau=-30}$ & $\bm{\tau=-40}$ \\
\Hline
\multirow{3}{*}{\shortstack{Resonant\\Vector Masses ($M_v^{(n)2}$)}} 
& 0     & 0     \\
& 13.89 & 18.90 \\
& 24.99 & 35.27 \\
& -     & 48.36 \\
\Hline
\multirow{3}{*}{$\mathcal{I}$} 
& \scalebox{0.85}{$\begin{pmatrix} 0 & 1.42 \\ -1.40 & 0 \\ 1.00 & 0\\ 0 & 1.00\end{pmatrix}$} 
& \scalebox{0.85}{$\begin{pmatrix} 0 & -0.29 & 0 \\ 0.29 & 0 & -2.41 \\ 0 & 2.41 & 0 \\ 1.00 & 0& 0\\ 0 & 1.00& 0\\ 0 & 0& 1.00\end{pmatrix}$} \\
\Hline
Rank of $\mathcal{I}$ & 2 & 3 \\
\Hline
Vector Singular Values $\sigma_n$ & (1.74, 1.72) & (2.63, 2.63, 1.00) \\
\Hline
\multirow{4}{*}{$\mathcal{M}^2$ Matrix} 
& \multirow{4}{*}{\scalebox{0.75}{$\begin{pmatrix} 1.00 & 0 & 0 & 1.42\\ 0 & 1.00 &-1.40 & 0\\ 0 &-1.40 & 13.89 & 0\\ 1.42 & 0 & 0 & 24.99\end{pmatrix}$}}
& \multirow{4}{*}{\scalebox{0.65}{$\begin{pmatrix} 1.00 & 0 & 0 & 0 & -0.29 & 0\\ 0 & 1.00 & 0 & 0.29 & 0 & -2.41\\ 0 & 0 & 1.00 & 0 & 2.41 & 0\\ 0 & 0.29 & 0 & 18.90 & 0 & 0\\ -0.29 & 0 & 2.41 & 0 & 35.27 & 0\\ 0 & -2.41 & 0 & 0 & 0 & 48.36\end{pmatrix}$}} \\
& & \\
& & \\
& & \\
\Hline
Type A Scalar Mass & (0.92, 25.07) & (0.83, 1.00, 35.44) \\
\Hline
Type B Scalar Mass & (0.85, 14.04) & (0.87, 18.90, 48.48) \\
\Hline
\end{NiceTabular}
\end{table}

This numerical analysis of the 6D braneworld model yields several critical insights that profoundly distinguish it from the 5D case:

\begin{itemize}
    \item \textbf{Full Rank and Vector Mode Correspondence:} Unlike the 5D scenario, where the skew-symmetry of the mixing matrix inevitably leads to zero singular values for odd-numbered bounded states, the vector KK modes in the 6D model couple simultaneously with two distinct channels of scalars corresponding to the two extra dimensions. This multi-channel coupling mechanism ensures that the mixing matrix $\mathcal{I}$ maintains full column rank; consequently, no zero singular values are observed in the vector sector. Physically, this implies that every bare vector state successfully absorbs a specific linear combination of scalars to acquire mass ($\sigma_a \neq 0$). As a result, the number of physical massive vector modes remains exactly equal to the number of resonant states derived from the bare eigenvalue equations.

    \item \textbf{Bifurcation and Survival of Massive Scalar Sectors:} The emergence of physical scalar degrees of freedom is rigorously governed by the interplay between the mixed absorption and the intrinsic scalar mass matrix $\mathcal{M}^2$. In the $d=2$ configuration, the vector modes absorb specific linear combinations of the scalar fields, defined by the column space of the full-rank mixing matrix $\mathcal{I}$. The true unabsorbed physical scalars reside in the orthogonal complement of these absorbed (Goldstone-like) directions. The physical identities and masses of these surviving modes are completely determined by the scalar mass matrix $\mathcal{M}^2$ in the effective action. 
    
Crucially, the symmetry properties of the integrands within the cross-coupling matrix $\widetilde{C}$ (e.g., the vanishing of overlap integrals for odd-parity integrands) enforce a block-diagonal structure on $\mathcal{M}^2$. Consequently, when projected onto the unabsorbed physical subspace, the surviving scalar spectrum naturally bifurcates into two decoupled sectors---Type A and Type B. Due to the intrinsic $C_{1,2}$ and $\widetilde{C}$ interactions, these remaining orthogonal combinations acquire substantial geometric masses (e.g., $0.92$ and $0.85$ at $\tau=-30$), rigorously establishing that higher-codimension geometries harbor an unavoidable and rich spectrum of massive physical scalar KK modes.
\end{itemize}

\section{Conclusion and Discussion}\label{discussions}

In this work, we have comprehensively investigated the mixing between the KK modes of a bulk $U(1)$ gauge field in braneworld scenarios with codimension $d$. We established that mixed vector-scalar and scalar-scalar couplings are generic and unavoidable features of these models, unless the extra-dimensional geometry satisfies the highly restrictive, separable condition $\partial_i \partial_j A = 0$. By utilizing the completeness of the basis functions and a well-defined Hilbert space inner product, we demonstrated that the gauge invariance of the resulting effective action is an intrinsic property, fundamentally independent of the choice of basis functions. This clarifies that previous geometric constraints on gauge symmetry found in the literature were largely artifacts of the stringent level-diagonal decoupling assumption.

Our analytical framework, supported by representative 5D and 6D numerical solutions, yields several key physical insights:

\begin{itemize}
    \item \textbf{Mass Shifts and EFT Truncation:} A primary finding is that the physical masses of the vector KK modes, identified algebraically as the singular values ($\sigma_a$) of the mixing matrix, deviate significantly from their unperturbed bare eigenvalues ($m_v^{(n)}$). Crucially, we demonstrate that the substantial reduction in the observable physical mass ($\sigma_a \ll m_v^{(n)}$) is an inherent consequence of the low-energy effective field theory (EFT) truncation. While incorporating the complete infinite scalar tower (including the continuum) would theoretically yield $\sigma_a = m_v^{(n)}$ via the completeness relation, any realistic truncation to a finite number of low-lying bound states intentionally breaks this exact completeness. Integrating out the inaccessible high-energy modes dynamically renormalizes the low-energy spectrum, manifesting macroscopically as a substantial downward shift in the observable vector KK masses.
    
    \item \textbf{Multi-channel Absorption and Survival of Massive Scalars}: In higher-codimension geometries ($d>1$), the existence of multiple scalar KK towers facilitates a multi-channel coupling mechanism. Unlike the 5D case where rank deficiency intrinsically leads to zero-singular-value residues, the multi-channel mixing matrix (e.g., in our 6D model) maintains full column rank, allowing every vector resonance to successfully absorb a specific linear combination of scalars. The true unabsorbed physical scalars reside precisely in the orthogonal complement to these absorbed directions. Crucially, the mass generation of these surviving modes is rigorously governed by the intrinsic scalar mass matrix $\mathcal{M}^2$. Driven by the parity-induced symmetries of the cross-coupling integrands, this unabsorbed scalar spectrum naturally bifurcates into decoupled physical sectors (e.g., Type A and Type B), acquiring substantial geometric masses. This establishes that higher-codimension braneworlds inevitably harbor a rich spectrum of massive physical scalar KK modes.

    \item \textbf{Symmetry-Induced Physical Scalar Spectra:} The emergence and mass generation of the unabsorbed physical scalars are rigorously governed by the intrinsic scalar mass matrix $\mathcal{M}^2$. The surviving physical scalars reside precisely in the orthogonal complement to the absorbed (Goldstone-like) directions. By projecting $\mathcal{M}^2$ onto this unabsorbed orthogonal subspace, the surviving scalar spectrum in our 6D model naturally bifurcates into decoupled physical sectors (Type A and Type B) due to the parity-induced block-diagonal structure of the overlap integrands. Driven by intrinsic scalar-scalar interactions, these surviving orthogonal combinations acquire substantial geometric masses, definitively proving that higher-codimension braneworlds naturally harbor a rich spectrum of massive physical scalar KK modes.

\end{itemize}

In conclusion, while the specific matrix representations depend on the parity and structure of the chosen wavefunctions, the underlying physical consequences---the dynamic renormalization of vector masses due to EFT truncation, and the unavoidable survival of massive orthogonal scalars---are robust features of higher-dimensional geometric mixing. Our results suggest that the interplay between extra-dimensional geometry, gauge symmetry, and multi-mode entanglement is far more complex than previously assumed, offering new theoretical avenues for exploring the dark sector, extended Higgs mechanisms, and the hierarchy problem in braneworld phenomenology.

%\bibliographystyle{JHEP}
%\bibliography{mixvsdof}

\providecommand{\href}[2]{#2}\begingroup\raggedright\endgroup

\end{document}